\documentclass[11pt, a4paper]{article}
\usepackage[textwidth = 7in, textheight=25cm,nohead]{geometry}
\usepackage{ulem}
\usepackage{graphicx, amssymb, amsmath, amsthm, amstext, booktabs, float, multirow, subfigure, rotating,natbib}
\usepackage{setspace}
\numberwithin{equation}{section} 
\numberwithin{figure}{section} 
\numberwithin{table}{section}

\theoremstyle{remark} 
\newtheorem{remark}{Remark}[section]

\bibpunct{(}{)}{,}{a}{,}{,}

\def\ppn{\vskip 6pt \noindent }
\def\R{{\mathbb{R}}}

\def\P{{\mathbb{P}}}
\def\E{{\mathbb{E}}}
\newcommand{{\Xs}}{{\cal X}}
\newcommand{{\Ys}}{{\cal Y}}
\newcommand{{\Ls}}{{\cal L}}
\newcommand{{\Ss}}{{\cal S}}
\newcommand{{\Gs}}{{\cal G}}
\newcommand{{\Hs}}{{\cal H}}
\newcommand{{\Ns}}{{\cal N}}
\newcommand{{\Is}}{{\cal I}}
\newcommand{{\Ks}}{{\cal K}}
\newcommand{{\Bs}}{{\cal B}}
\newcommand{{\Cs}}{{\cal C}}
\newcommand{{\Rs}}{{\cal R}}
\newcommand{{\Us}}{{\cal U}}
\newcommand{{\Ds}}{{\cal D}}
\newcommand{{\Fs}}{{\cal F}}

\newcommand{{\pp}}{{\mathbf p}}
\newcommand{{\KK}}{{\mathbf K}}
\newcommand{{\HH}}{{\mathbf H}}
\newcommand{{\II}}{{\mathbf I}}
\newcommand{{\ZZ}}{{\mathbf Z}}
\newcommand{{\yy}}{{\mathbf y}}
\newcommand{{\ab}}{{\mathbf a}}
\newcommand{{\zz}}{{\mathbf z}}

\newcommand{{\toP}}{{\overset{P}{\longrightarrow}\ }}

\newcommand{{\toL}}{{\overset{\mathcal{L}}{\longrightarrow}\ }}

\newcommand{{\dou}}{$\leadsto$\ }

\DeclareMathOperator{\VaR}{VaR}

\newcommand{\indic}[1]{
\hbox{${\it 1}\hskip -4.5pt I_{\{ #1 \}}$}
}

\begin{document}

\setlength{\belowdisplayskip}{5pt} \setlength{\belowdisplayshortskip}{3pt}
\setlength{\abovedisplayskip}{5pt} \setlength{\abovedisplayshortskip}{0pt}

\title{A nonparametric copula approach to conditional Value-at-Risk }
\author{\sc{Gery Geenens}\thanks{Corresponding author: ggeenens@unsw.edu.au, School of Mathematics and Statistics, UNSW Sydney, Australia, tel +61 2 938 57032, fax +61 2 9385 7123 } \\School of Mathematics and Statistics,\\ UNSW Sydney, Australia  \and \sc{Richard Dunn}\\School of Mathematics and Statistics,\\ UNSW Sydney, Australia }
\date{}
\maketitle
\thispagestyle{empty} 

\begin{abstract}
Value-at-Risk and its conditional allegory, which takes into account the available information about the economic environment, form the centrepiece of the Basel framework for the evaluation of market risk in the banking sector. In this paper, a new nonparametric framework for estimating this conditional Value-at-Risk is presented. A nonparametric approach is particularly pertinent as the traditionally used parametric distributions have been shown to be insufficiently robust and flexible in most of the equity-return data sets observed in practice. The method extracts the quantile of the conditional distribution of interest, whose estimation is based on a novel estimator of the density of the copula describing the dynamic dependence observed in the series of returns. Monte-Carlo simulations and real-world back-testing analyses demonstrate the potential of the approach, whose performance may be superior to its industry counterparts.
\end{abstract}

\section{Introduction}\label{sec:intro}

\ppn In the past two decades the quantification of risk has become an area of high interest due to its paramount role in modern financial sectors, as core business operations are predicated on mutually beneficial trades of this risk \citep{Segal11}. While there are numerous methodologies for quantifying risk, few are as popular and widespread as Value-at-Risk (hereafter: VaR). VaR became a crucial means for financial risk management after the stock market crash of 1987, and it is now globally accepted as benchmark for risk management through its inclusion in the mandatory Basel II Banking standard in 2004. Practically, VaR is the amount of capital that a firm has to secure aside to resist unlikely but not impossible adverse events when engaging in risky trading activities. See \cite{Duffie97,Jorion01} and \cite{Scaillet03} for the financial background and applications. Statistically speaking, VaR is nothing but an upper quantile of the distribution $F_X$ of some random loss $X$, potentially faced by the firm over a given period when engaging in those activities. For a certain fixed level $\alpha \in (0,1)$, with $\alpha$ generally close to 1, VaR is thus defined as:
\[\VaR_\alpha(X) = \inf \{ x \in   \mathbb{R}: \P(X > x) \leq 1-\alpha \} = F_X^{-1}(\alpha), \]
where $F_X^{-1}(\alpha)$ is the (generalised) inverse of $F_X$, i.e., its quantile function.

\ppn It is, however, clear that the risk that some trading activities represents varies with the market conditions. Hence, it is generally paramount to assess that risk {\it conditionally} on some additional variables, say $\ZZ \in \R^d$, reflecting the latest available information about the economic environment \citep{Chernozhukov01,McNeil05,Kuester06,Escanciano10}. In a general framework, the conditional Value-at-Risk ($c$VaR) at level $\alpha \in (0,1)$ for a random loss $X$ given some vector of covariates $\ZZ$, is defined as 
\begin{equation} c\VaR_\alpha(X|\ZZ = \zz)=\inf \{ x \in   \mathbb{R}: \P(X > x|\ZZ=\zz) \leq 1-\alpha \} = F_{X|\ZZ}^{-1}(\alpha|\zz), \label{eqn:$c$VaR}\end{equation}
that is, the $\alpha$-quantile of the conditional distribution $F_{X|\ZZ}$ of $X$ given $\ZZ=\zz$. Among others, the variables $\ZZ$ may be past observed losses, in a time series setting (see Section \ref{sec:framework}), and/or other exogenous economic and market covariates. 

\begin{remark} The idiom `conditional Value-at-Risk' has sometimes been used differently in the previous literature, e.g.\ in \cite{Jorion01} or \cite{Rockafellar02}, for what is now more commonly called the Expected Shortfall (ES). That should not bring any confusion here. Some comments about the related problem of estimating the conditional ES are provided in Section \ref{sec:ccl}.
\end{remark}

\ppn Attempts to estimate $c$VaR have historically centred on parametric models due to their familiar and convenient nature. The most well-known of these approaches is probably the industry-benchmark RiskMetrics \citep{Morgan96}, which attempts to characterise future returns through a normal distribution that is scaled by an Exponentially Weighted Moving Average (EWMA) estimate of the market volatility. This is despite empirical research revealing that the normal assumption usually falls short for fitting financial data, which typically show substantial skewness and kurtosis. This has motivated the development of EWMA-models replacing the normal innovations with ones from the $t$-distribution \citep{So06}, the Laplace distribution \citep{Guermat01}, the asymmetric-Laplace distribution \citep{Gerlach13}, or even more sophisticated parametric distributions. In spite of this, RiskMetrics has arguably remained the industry standard since 1996.

\ppn RiskMetrics is based on modelling the volatility embedded in the series of returns via an integrated GARCH model with Normal innovations. This belongs to a much wider class of parametric time series models, which can be specified as follows. Let $X_t$ be the return (or loss) of interest at time $t$. Then, for $t=0,\ldots,T$, it is assumed that
\begin{align}
X_{t+1} | \Fs_t \quad & \sim \quad \Ds(\xi_{t},\theta), \label{eqn:GGARCH1}\\
\xi_{t+1} & = h(\xi_{t},\ldots,\xi_{t-p+1},X_{t},\ldots,X_{t-q+1};\beta), \label{eqn:GGARCH2}
\end{align}
where $\Fs_t$ is the information gathered from time 0 up to time $t$, $\Ds$ is a specified parametric innovation distribution which is parametrised by $\xi_t$, a vector of time-varying parameters, and $\theta$, a vector of static parameters, and $h$ is the function, parametrised by a static vector of parameters $\beta$, which sets the dynamics of the considered model. For instance, a popular approach for modelling the behaviour of volatility (i.e \ the conditional variance of $X_t$), is through the GARCH$(1,1)$ model \citep{Bollerslev86}, which satisfies the above specification with $\xi_t = \sigma_t^2$, the volatility parameter at time $t$, and $h$ given by $\sigma_{t+1}^2 = \beta_0 + \beta_1 \epsilon^2_{t} + \beta_2 \sigma^2_{t}$. The conditional distribution $\Ds$ is usually assumed to be Normal, Student or skew-Student. 
Further specifications of GARCH-like models satisfying dynamics like (\ref{eqn:GGARCH1})-(\ref{eqn:GGARCH2}) include the above-mentioned integrated GARCH (iGARCH) model \citep{Engle86},  Exponential GARCH model (eGARCH) \citep{Nelson91} and the Asymmetric Power ARCH model (APARCH) \citep{Ding93}, with the gjr-GARCH specification of \cite{Glosten93} as special case. For a recent and extensive survey of GARCH models, see \cite{Francq11}. Another class of models satisfying (\ref{eqn:GGARCH1})-(\ref{eqn:GGARCH2}) are the Generalised Autoregressive Score models (GAS) recently suggested in \cite{Creal13} and \cite{Harvey13}, which seem to be serious alternatives to GARCH models when modelling highly non-linear volatility dynamics. Alternative parametric models for forecasting $c$VaR finally include approaches based on Extreme-Value-Theory, such as Block Maxima Model (BMM, \citet{McNeil98,McNeil99}) or Peak over Threshold models (POT, \citet{Embrechts97,Embrechts99,Chavez14}); or on quantile regression such as CaViaR \citep{Engle06}. \cite{Nieto16} offers a recent comprehensive review.

\ppn While those approaches have their merits, they typically suffer from the usual rigidity of the parametric setting. In particular, different parametric model specifications usually lead to disparate results which are hard to reconciliate, see empirical illustration of this in \citet[Section 4.2]{Li08}. In addition, model misspecification can have serious consequences; e.g., it has been argued that the 2009 global financial crisis was mainly due to an unwarranted usage of the parametric Gaussian copula model for asset pricing \citep{Salmon09}. In answer to this lack of flexibility, nonparametric models, which discard any rigid structure for the data, have been suggested as well. The nonparametric methodology really lets the data `speak for itself', hence its attractiveness. This is particularly important when estimating high quantiles of distributions, such as VaR, as usually the tail behaviour of a probability density is entirely locked by its parametric specification. On the other hand, estimation of tails of distributions without any parametric guidelines is usually a challenging task, owing to the typical sparseness of data in those areas.

\ppn First nonparametric attempts centred on empirical methodologies in which past returns were assumed to represent the full distribution of future returns, see e.g.\ the historical simulation technique in \citet{Hendricks96,Linsmeier00,Dowd01} and \citet{Chen07}. This basic approach ignores the changing nature of equity markets, hence \cite{Barone-Adesi02}, following \cite{Hull98}, suggested a scaling of the historical returns by the implied market volatility in order to compensate for these changes. This method, known as Filtered Historical Simulation (FHS) has grown to be a forerunner in the market due to its simplicity and computational efficiency. More recently, \cite{Zikovic11} and \citet{Dupuis15} proposed `weighted' versions of historical simulations along similar lines. \cite{Fan03} suggested what can be regarded as a semiparametric version of RiskMetrics, later extended in \citet{Martins06,Martins16} and \cite{Wang16}. 

\ppn \cite{Cai02} used a Nadaraya-Watson-type estimator of the conditional distribution of $X$ given $\ZZ$ and obtained the conditional VaR by inverting it, an approach refined in \cite{Scaillet05,Cai08,Li08,Taylor08,Wu08,Xu13} and \cite{Franke15}. Some of these estimators are discussed in more detail in Section \ref{subsec:kcde}, 
as this paper actually gives a continuation to them in that it studies a novel estimator of $c\VaR_\alpha(X|\ZZ = \zz)$ based on the direct inversion of a nonparametric estimator of the conditional distribution $F_{X|\ZZ}$. What mostly differs from the previous contributions is that here $F_{X|\ZZ}$ will be estimated by making use of new developments in the field of {\it nonparametric copula modelling}. This has various advantages, those being detailed throughout the paper. Of course, copulas have been around for a long time in finance and related fields, see e.g.\ \cite{Embrechts02,Embrechts09,Cherubini04,Cherubini12} for comprehensive reviews, but the literature in the field is again overwhelmingly dominated by parametric methods. Through the particular application of estimating $c$VaR, this paper aims to demonstrate the capability of flexible nonparametric copula modelling methods for financial applications.

\ppn It is organised as follows: Section \ref{sec:framework} sets the framework that will be considered. Section \ref{subsec:kcde} provides a short review of existing nonparametric, kernel-based, methods for estimating conditional distributions and densities which this work will complement. Section \ref{sec:est} outlines, in detail, how some recent elements of nonparametric copula modelling can be amalgamated to provide an estimate of the conditional Value-at-Risk. Section \ref{sec:realdat} focuses on empirically illustrating and validating the idea through a Monte-Carlo simulation study and two real data applications. Section \ref{sec:ccl} concludes with some paths for future research.

\section{Framework} \label{sec:framework}

Consider a sample $\{X_t;t=0,1,\ldots,T-1\}$ of losses\footnote{Here, the term `loss' for $X$ is quite generic. It may mean negative log-returns, for instance, or any other quantity that could be appropriate.} for a given portfolio, and assume that it is a realisation of a strictly stationary\footnote{Within the considered nonparametric framework, the strict stationarity of $\Xs$ is usually assumed \citep{Scaillet05,Cai08,Franke15}. The reason is that nonparametric procedures are explicitly based on empirical estimation of the conditional distribution (\ref{eqn:condistr}). Hence this distribution must be identifiable from the observed sample, which is out of reach under a weak stationarity assumption on $\Xs$.} process $\Xs$ in discrete time, with marginal distribution $F_X$ admitting a density $f_X$. The theoretical considerations exposed in Section \ref{subsec:estimation} hold assuming that $\{X_t\}$ forms an $\alpha$-mixing (i.e., strongly mixing) sequence, a dependence structure obeyed by most time series models \citep{Doukhan94}, hence are rather general. 

\ppn For simplicity, this paper will only focus on the problem of estimating the {\it one-step-ahead} $c$VaR, that is, estimating at time $T-1$ the VaR at time $T$; when taking as `influencing economic factors', i.e.\ $\ZZ$ in (\ref{eqn:$c$VaR}), the just observed loss $X_{T-1}$ only. Specifically, the parameter of interest is here
\[c\VaR_\alpha(X_T|X_{T-1} = x)=\inf \{ y \in   \mathbb{R}: \P(X_T > y|X_{T-1}=x ) \leq 1-\alpha \} = F^{-1}_{X_T|X_{T-1}}(\alpha|x), \] 
the quantile of level $\alpha$ of the conditional distribution of $X_T$ given that $X_{T-1} = x$, viz.
\begin{equation} F_{X_T|X_{T-1}}(y|x) = \P(X_T \leq y | X_{T-1} = x). \label{eqn:condistr} \end{equation}
Although very basic, this Markov-type framework provides scope for the volatility clustering and serial correlation of returns seen in real datasets to be taken into account \citep{McNeil05}. In addition, it covers the effect of return momentum \citep{Cahart97}, which has been demonstrated to have the strongest effect on future equity returns compared to any of the other popular risk factors \citep{Bender13}. Finally, it is consistent with \cite{Fama65}'s {\it Efficient Market Hypothesis}. Hence it seems a valid basis for illustrating the idea.

\begin{remark} It is stressed, though, that the suggested methodology can readily be extended to more general frameworks. In particular, conditioning on more than one variable would be conceptually straightforward, bearing in my mind that, the developed procedure being purely nonparametric, it could suffer from dimensionality issues \citep{Geenens11}. If need be, dimension reduction can be achieved by introducing some structural assumptions, such as a Single-Index structure, see \cite{Fan16}.
\end{remark}

\ppn 
Interestingly, $F_{X_T|X_{T-1}}(y|x)$ can be regarded as a regression function, arguing that 
\begin{equation} F_{X_T|X_{T-1}}(y|x) = \E(\indic{X_T \leq y}|X_{T-1}=x), \label{eqn:condistrreg} \end{equation}
where $\indic{\cdot}$ is the indicator function, equal to 1 if the statement between brackets is true and 0 otherwise. Naturally, it can also be written 
\begin{equation} F_{X_T|X_{T-1}}(y|x) = \int_{-\infty}^{y} f_{X_T|X_{T-1}}(\xi|x)\,d\xi, \label{eqn:conddistr} \end{equation}
where $f_{X_T|X_{T-1}}$ is the conditional density of $X_T$ given $X_{T-1}$ -- provided it exists. One can, therefore, estimate $F_{X_T|X_{T-1}}$ using either regression ideas, or by plugging an estimate of the conditional density, say $\hat{f}_{X_T|X_{T-1}}$, in (\ref{eqn:conddistr}). Regression-based estimation has been quite popular (see next section), however, an estimator of type
\begin{equation} \hat{F}_{X_T|X_{T-1}}(y|x) = \int_{-\infty}^{y} \hat{f}_{X_T|X_{T-1}}(\xi|x)\,d\xi \label{eqn:estconddistr} \end{equation}
offers substantial advantages. Indeed, provided that $\hat{f}_{X_T|X_{T-1}}$ is a {\it bona fide} density, in the sense that $\hat{f}_{X_T|X_{T-1}}(y|x) \geq 0$ $\forall (x,y)$ and $\int_{-\infty}^\infty \hat{f}_{X_T|X_{T-1}}(\xi|x)\,d\xi =1$ $\forall x$, the so-obtained $ \hat{F}_{X_T|X_{T-1}}$ is always a {\it bona fide} distribution function as well, non-decreasing in $y$ and with $\hat{F}_{X_T|X_{T-1}}(-\infty|x) = 0$, $\hat{F}_{X_T|X_{T-1}}(\infty|x) = 1$. This is essential when inverting it for obtaining its quantile. Regression-based approaches may lead to estimates not constrained to lie in $[0,1]$ or to be monotonic in $y$, which causes obvious issues and inconsistencies.

\section{Kernel estimators of conditional distributions and densities} \label{subsec:kcde}

The most basic nonparametric regression estimator is arguably the Nadaraya-Watson (NW) estimator \citep{Nadaraya64,Watson64}, which for (\ref{eqn:condistrreg}) writes
\begin{equation} \tilde{F}_{X_T|X_{T-1}}(y|x) = \frac{\sum_{t=1}^{T-1} K_h(x-X_{t-1})\indic{X_t \leq y}}{\sum_{t=1}^{T-1} K_h(x-X_{t-1})}, \label{eqn:NWconddistr} \end{equation} 
where $K$ is a symmetric probability distribution (`kernel'), $h>0$ is a smoothing parameter (`bandwidth') and $K_h(\cdot) = K(\cdot/h)/h$, see \citet[Chapter 4]{Hardle04} for details. Clearly, the so-defined $\hat{F}_{X_T|X_{T-1}}(y|x)$ is a {\it bona fide} distribution function, always lying in $[0,1]$ and non-decreasing in $y$. This makes the inversion of (\ref{eqn:NWconddistr}) very easy, and \cite{Franke15} studied in detail the estimator of the conditional VaR obtained by doing so. 

\ppn It is usually accepted, though, that local polynomial regression estimators \citep{Fan96} enjoy better theoretical properties than the Nadaraya-Watson estimator. This motivated \cite{Yu98} to suggest a local linear (LL) estimator of a conditional distribution function. Yet, the LL estimator is neither constrained to between 0 and 1, nor to be monotonic in $y$, which violates common sense. Hence \cite{Hall99} and \cite{Cai02} proposed the weighted Nadaraya-Watson estimator (WNW), which satisfies those constraints while sharing the same theoretical properties as the LL estimator. However, this estimator is not continuous in its first argument, which may not be ideal. In addition, for each $x$, it requires numerically determining a set of weights through a constrained optimisation problem, which may be computationally demanding. Note that \cite{Cai08} fixed the non-continuity issue through yet another layer of smoothing, suggesting their `{\it weighted double kernel local linear estimator}' (WDKLL) of $F_{X_T|X_{T-1}}$, which nonetheless still necessitates numerical optimisation for determining the right weights, making the procedure rather cumbersome. An approach based on the conditional density through (\ref{eqn:conddistr}) may be simpler. 

\ppn By definition, the conditional density ${f}_{X_T|X_{T-1}}$ is
\begin{equation}  f_{X_T|X_{T-1}}(y|x) = \frac{f_{X_{T-1},X_T}(x,y)}{f_{X_{T-1}}(x)}, \label{eqn:defconddens} \end{equation}
where $f_{X_{T-1},X_T}$ is the joint density of the vector $(X_{T-1},X_T)$ and $f_{X_{T-1}}$ its marginal. Again, these quantities are independent of $T$, by the assumed stationarity of the process $\Xs$, hence $f_{X_{T-1},X_T} \doteq f_1$, say, and $f_{X_{T-1}} = f_X$.  This motivates a nonparametric estimator of type 
\begin{equation} \hat{f}_{X_T|X_{T-1}}(y|x) = \frac{\hat{f}_1(x,y)}{\hat{f}_{X}(x)}, \label{eqn:conddens} \end{equation}
where both the numerator and denominator in (\ref{eqn:defconddens}) are estimated from the observed sample by usual kernel-type estimators $\hat{f}_{1}$ and $\hat{f}_{X}$ for bivariate and univariate densities \citep[Chapter 3]{Hardle04}. This kind of `plug-in' kernel conditional density estimator, initially proposed in \cite{Rosenblatt68}, was studied in \cite{Hyndman96,Bashtannyk01,Fan04} and \cite{Hall04}, among others. Plugging (\ref{eqn:conddens}) into (\ref{eqn:conddistr}) yields the `double kernel' Nadaraya-Watson estimator (DKNW):
\begin{equation} \hat{F}_{X_T|X_{T-1}}(y|x) = \frac{\sum_{t=1}^{T-1} K_h(x-X_{t-1})\Ks_0((y-X_{t})/{h_0})}{\sum_{t=1}^{T-1} K_h(x-X_{t-1})}, \label{eqn:DKNWconddistr} \end{equation} 
where $\Ks_0(u) = \int_{-\infty}^u K_0(u^*)\,du^*$ is the `integrated' version of a kernel $K_0$ and $h_0$ is a bandwidth. Essentially, (\ref{eqn:DKNWconddistr}) is (\ref{eqn:NWconddistr}) but with the indicator $\indic{X_t \leq y}$ replaced by a smoothed version of it, which makes $\hat{F}_{X_T|X_{T-1}}(y|x)$ continuous in $y$. This is usually beneficial to the estimator, as it has often been stressed in the classical literature on quantile estimation \citep{Azzalini81,Falk85,Yang85,Bolance14}. It is precisely this estimator (\ref{eqn:DKNWconddistr}) that \cite{Scaillet05} and \cite{Li08} inverted to produce their estimator of $c$VaR -- see also \cite{Ferraty16} for an extension of this to a functional context. In the study below, it will therefore be taken as the benchmark for the previously proposed kernel-type methods. 

\ppn Yet, it can be understood that the ratio form of (\ref{eqn:conddens}) creates issues \citep{Faugeras09}, both in theory and in practice: denominator close to 0, numerical instability, delicate choice of smoothing parameters, etc. What is suggested in this paper is to use a different type of kernel estimator of the conditional density in (\ref{eqn:estconddistr}), based on the copula of the vector $(X_{T-1},X_T)$. Given that $X_{T-1}$ and $X_T$ have the same distribution by stationarity, the only extra information contained in the joint distribution of $(X_{T-1},X_T)$ relates to their dependence. This evidences the appropriateness of an approach based on copulas, a.k.a.\ dependence functions, in this framework.

\section{Nonparametric copula-based estimation of $c$VaR}  \label{sec:est}

\subsection{Copulas, copula densities, conditional densities and conditional distributions}

The central result of copula theory, known as Sklar's theorem \citep{Sklar59}, asserts that for any continuous bivariate random vector $(X,Y)$ with distribution function $F_{XY}$ (and marginals $F_X$ and $F_Y$), there exists a unique `copula' function $C$ such that
\begin{equation} F_{XY}(x,y) = C(F_X(x),F_Y(y)) \qquad \forall (x,y) \in \R^2.  \label{eqn:copdef} \end{equation}
Clearly, $C$ describes how $X$ and $Y$ `interact' to produce the joint behaviour of $(X,Y)$, hence it fully characterises the dependence structure between $X$ and $Y$ and isolates it from their marginal behaviours. See \cite{Joe97} and \cite{Nelsen06} for general textbook treatment of these ideas, and \cite{Cherubini04,Cherubini12} for their implications in finance. Importantly, given that $F_X(X), F_Y(Y) \sim \Us_{[0,1]}$ (probability integral transform), $C$ is actually a bivariate distribution on the unit square $\Is \doteq [0,1]^2$ with uniform marginals. Under mild conditions, that distribution admits a density, known as the {\it copula density}:
\[c(u,v) = \frac{\partial^2 C}{\partial u \partial v}(u,v), \quad (u,v) \in \Is. \]

\ppn Writing (\ref{eqn:copdef}) for the vector $(X_{T-1},X_T)$ yields
\[F_{X_{T-1},X_T}(x,y) = C_1(F_{X}(x),F_{X}(y)), \]
for some copula $C_1$ independent of $T$ and $F_X = F_{X_{T-1}} = F_{X_T}$ by stationarity. Now, differentiating both sides, the joint density $f_1=f_{X_{T-1},X_T}$ is seen to be
\[f_1(x,y) =\frac{\partial^2 F_{X_{T-1},X_T}}{\partial x \partial y}(x,y) = c_1(F_{X}(x),F_{X}(y))f_{X}(x)f_{X}(y),\]
by the chain rule, with $c_1$ the copula density of $C_1$. Hence, from (\ref{eqn:defconddens}), the conditional density of $X_T$ given $X_{T-1}$ can be written directly in terms of the copula density:
\begin{equation} f_{X_T|X_{T-1}}(y|x) =   c_1(F_{X}(x),F_{X}(y))f_{X}(y). \label{eqn:conddenscop} \end{equation}
Replacing the unknown $c_1$, $F_X$ and $f_X$ by some estimators $\hat{c}_1$, $\hat{F}_X$ and $\hat{f}_X$ leads to a different, copula-based estimator of $f_{X_T|X_{T-1}}$, viz.
\begin{equation} \tilde{f}_{X_T|X_{T-1}}(y|x) =   \hat{c}_1(\hat{F}_{X}(x),\hat{F}_{X}(y))\hat{f}_{X}(y). \label{eqn:estconddenscop} \end{equation}
Under a product shape, this estimator is, unlike (\ref{eqn:conddens}), free from any issues a random denominator creates, hence its attractiveness. Expression (\ref{eqn:conddenscop}) is also very intuitive: the conditional density of $X_T$ given $X_{T-1}$ is the marginal density of $X_T$, corrected for the influence that $X_{T-1}$ may have on $X_T$ through the copula density $c_1$ of the vector $(X_{T-1},X_T)$. Now, integrating (\ref{eqn:conddenscop}) in (\ref{eqn:conddistr}) yields
\begin{equation*} F_{X_T|X_{T-1}}(y|x)  = \int_{-\infty}^{y} c_1(F_{X}(x),F_{X}(\xi))f_{X}(\xi)\,d\xi = \int_{0}^{F_X(y)} c_1(F_{X}(x),v)\,dv, \end{equation*}
through the change-of-variable $F_X(\xi) \doteq v$. Such an expression is known in the copula literature as a $h$-function. 

\subsection{Nonparametric copula density estimation and $c$VaR} \label{subsec:estimation}

Usually, $h$-functions are estimated by numerically integrating an estimator of $c_1$ \citep{Nagler15}. However, seeing that 
\[F_{X_T|X_{T-1}}(y|x) = \E_X\left(c_1(F_X(x),F_X(X))\indic{X\leq y} \right),\]
the integral can easily be approximated by Monte-Carlo, viz.
\begin{equation*} \hat{F}_{X_T|X_{T-1}}(y|x) = \frac{1}{T} \sum_{t=0}^{T-1} \hat{c}_1(\hat{F}_{X}(x),\hat{F}_{X}(X_t))\indic{X_t \leq y}, \end{equation*} 
where $\hat{c}_1$ and $\hat{F}_X$ are appropriate estimators of the unknown $c_1$ and $F_{X}$ as in (\ref{eqn:estconddenscop}) (but here the estimation of the marginal density $f_X$ is not required). Interestingly, estimator (\ref{eqn:estconddistrcop}) can be thought of as a weighted empirical distribution function, where each indicator $\indic{X_t \leq y}$ is weighted according to the chance of seeing $X_t$ if it is known that $X_{t-1} = x$, as measured by the (estimated) copula density $\hat{c}_1$. Note that, for some reasons briefly mentioned earlier, one might prefer the continuous version
\begin{equation} \hat{F}_{X_T|X_{T-1}}(y|x) = \frac{1}{T} \sum_{t=0}^{T-1} \hat{c}_1(\hat{F}_{X}(x),\hat{F}_{X}(X_t))\Ks_0((y-X_{t})/{h_0}), \label{eqn:estconddistrcop} \end{equation} 
where $\Ks_0$ and $h_0$ are as in (\ref{eqn:DKNWconddistr}).

\ppn In the copula framework, it is customary to estimate $F_X$ by the rescaled empirical distribution function
\[\hat{F}_X(x) = \frac{1}{T+1} \sum_{t=0}^{T-1} \indic{X_t \leq x}, \]
known to be a simple and uniformly consistent estimator of $F_X$. It is, therefore, clear that the viability of estimator (\ref{eqn:estconddistrcop}) is mostly conditional on a reliable estimator for $c_1$. However, good nonparametric estimation of a copula density has long proved elusive. This mainly for three reasons which make the case non-standard: 1) a copula density has bounded support $\Is$, and kernel estimators are known to heavily suffer from boundary bias issues; 2) a copula density is often unbounded in some corners of $\Is$, and kernel estimators are known not to be consistent in that case; and 3) there is no access to genuine observations from the density to estimate, as $c_1$ is essentially the density of $(F_X(X_{T-1}),F_X(X_{T}))$ where $F_X$ is unknown, and one must resort to `pseudo-observations' $\{\hat{F}_X(X_t)\}$.

\ppn Recently, though, \cite{Geenens16}, after an original idea of \cite{Geenens14a}, proposed a novel kernel-type estimator of the copula density along the following lines. Define 
\[S_{T-1} = \Phi^{-1}(F_X(X_{T-1})) \quad \text{ and } \quad S_{T} = \Phi^{-1}(F_X(X_{T})), \]
where $\Phi^{-1}$ is the `probit' function, i.e.\ the inverse of the standard normal distribution function $\Phi$. Through standard distributional arguments, one gets 
\[c_1(u,v) = \frac{g_{1}(\Phi^{-1}(u),\Phi^{-1}(v))}{\phi(\Phi^{-1}(u))\phi(\Phi^{-1}(v))},\qquad \forall (u,v) \in \Is, \]
where $g_1$ is the joint density of $(S_{T-1},S_T)$ and $\phi$ is the standard normal density. Hence an estimator $\hat{c}_1$ of $c_1$ can be directly obtained from any estimator $\hat{g}_1$ of $g_1$:
\begin{equation} \hat{c}_1(u,v) = \frac{\hat{g}_{1}(\Phi^{-1}(u),\Phi^{-1}(v))}{\phi(\Phi^{-1}(u))\phi(\Phi^{-1}(v))}. \label{eqn:chat}  \end{equation}
Estimating $g_1$, though, is much easier. For $U \sim \Us_{[0,1]}$, $\Phi^{-1}(U) \sim \Ns(0,1)$, hence $g_1$ has unconstrained support with standard normal marginals. Local likelihood methods, in particular \cite{Loader96}'s local log-quadratic estimator, are particularly good at estimating normal densities, hence the appropriateness of using this type of methodology for estimating $g_1$ in this context. 

\ppn \cite{Geenens16}'s estimator, called `LLTKDE2', is actually (\ref{eqn:chat}) with $\hat{g}_1$ being the local log-quadratic estimator of the bivariate density $g_1$ based on pseudo-observations $\{\Phi^{-1}\left(\hat{F}_X(X_t) \right)\}$. Its theoretical properties were obtained. In particular, under mild assumptions, it was shown to be uniformly consistent on any compact proper subset of $\Is$, and asymptotically normal with known expressions of (asymptotic) bias and variance. In addition, a practical criterion for selecting the always crucial smoothing parameters was studied and tested. Combining transformation and local likelihood estimation, the procedure actually takes advantage of the known uniform margins of $C_1$, which results in remarkably accurate estimation \citep{Geenens16,Nagler15,DeBacker16}. Besides, the LLTKDE2 estimates typically enjoy a visually pleasant appearance usually peculiar to parametric fits. 

\ppn What is suggested in this paper is to use that LLTKDE2 estimator $\hat{c}_1$ in (\ref{eqn:estconddistrcop}) and proceed with the extraction of $c$VaR. The nonparametric copula-based $c$VaR estimator is thus defined as 
\[c\widehat{\VaR}_{\alpha,T}(x) \doteq c\widehat{\VaR}_\alpha(X_T|X_{T-1}= x)  = \hat{F}^{-1}_{X_T|X_{T-1}}(\alpha|x), \]
where $\hat{F}^{-1}_{X_T|X_{T-1}}$ is the generalised inverse of (\ref{eqn:estconddistrcop}). Given that $\hat{F}_X$ is uniformly consistent for $F_X$ on $\R$ and $\hat{c}_1$ is uniformly consistent for the integrable $c_1$ on any proper compact subset of $\Is$, (\ref{eqn:estconddistrcop}) is also uniformly consistent over any compact subset of $\R^2$ by the ergodic theorem. It classically follows that, provided that $\inf_{x \in G} f_{X_T|X_{T-1}}(c\VaR_{\alpha,T}(x)|x) > 0$ where $G$ is any compact subset of $\R$, $c\widehat{\VaR}_{\alpha,T}(x)$ is a uniformly consistent estimator of $c\VaR_{\alpha,T}(x) \doteq c\VaR_{\alpha}(X_T|X_{T-1}=x)$:
\[\sup_{x \in G} |c\widehat{\VaR}_{\alpha,T}(x)-c\VaR_{\alpha,T}(x)| \toP 0 \qquad \text{ as } T \to \infty, \]
for any fixed $\alpha \in (0,1)$. Further theoretical properties (such as asymptotic normality and rate of convergence) of this estimator would easily follow from the usual asymptotic representation of sample quantiles, viz.
\[c\widehat{\VaR}_{\alpha,T}(x)-c\VaR_{\alpha,T}(x) \simeq  \frac{\alpha - \hat{F}_{X_T|X_{T-1}}(c\VaR_{\alpha,T}(x)|x)}{f_{X_T|X_{T-1}}(c\VaR_{\alpha,T}(x)|x)},\]
but details are left aside owing to the rather unwieldy expressions in \cite{Geenens16}. Rather, its practical performance is assessed in Section \ref{sec:realdat} via real-data analyses, Monte-Carlo simulations and back-testing.

\section{Empirical study} \label{sec:realdat}

\subsection{Illustration - S$\&$P500 data} \label{subsec:SNP}

\begin{figure}[H] \centering
\includegraphics[width=.7\textwidth]{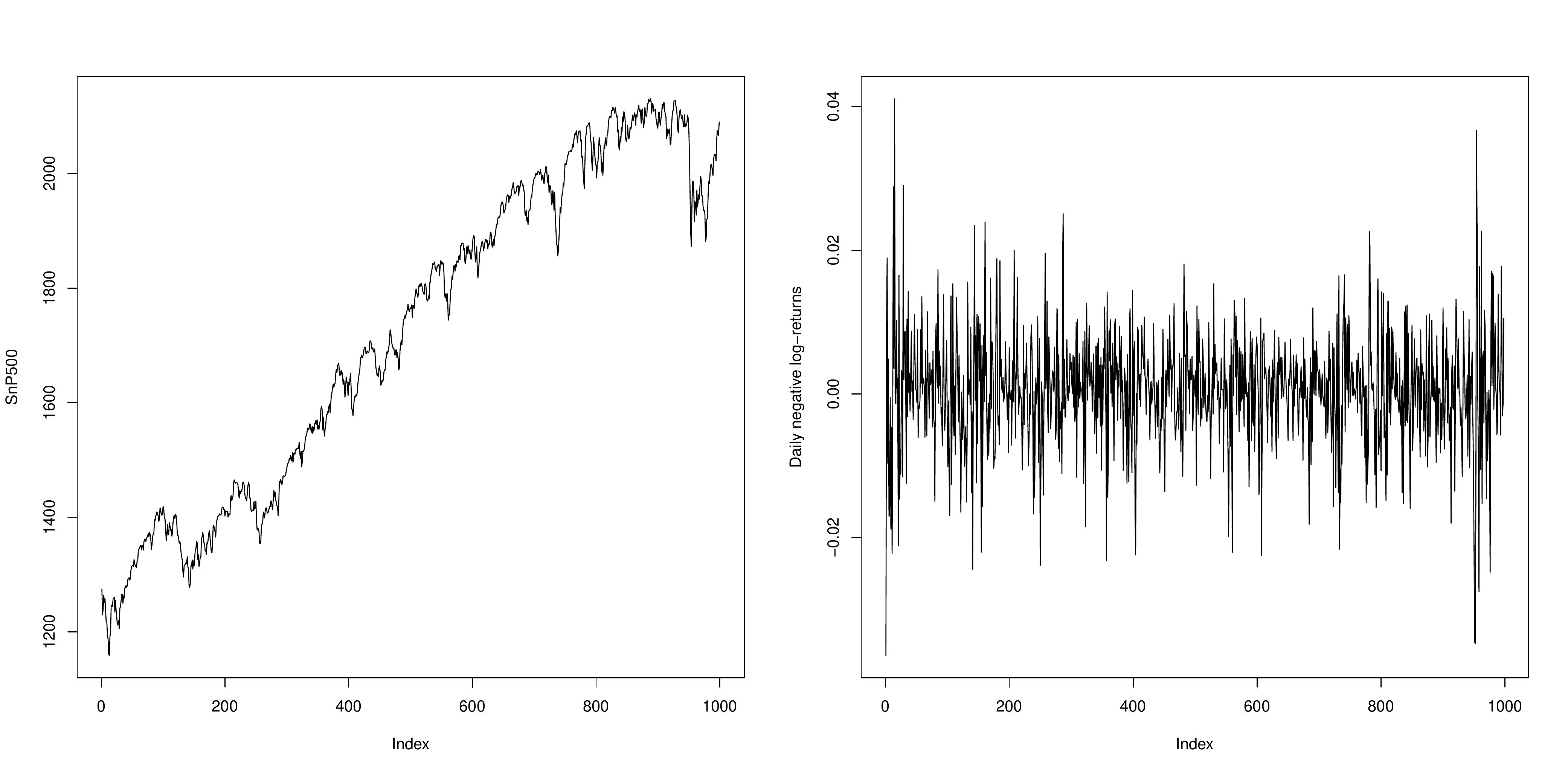}
\caption{S$\&$P500 index (left) and corresponding negative log-return series (right) for the period November 2011 to October 2015 (1,000 trading days).}
\label{fig:data}
\end{figure}

Firstly the procedure described in the previous sections is illustrated on the daily returns from the S$\&$P500 US index between November 2011 and October 2015. 
The S$\&$P500 index is one of the most actively traded indices available to investors. Figure \ref{fig:data} (left) shows the evolution over time of the S$\&$P500 index from the 9th of November 2011 until the 31st of October 2015 (which corresponds to $T=$ 1,000 trading days). Figure \ref{fig:data} (right) shows the corresponding negative log-returns series $\{X_t\}$, i.e., approximately the percentage changes in the value of the index. It is reasonable to posit that this series is (at least approximately) stationary. 

\begin{figure}[H] \centering
\includegraphics[width=.45\textwidth]{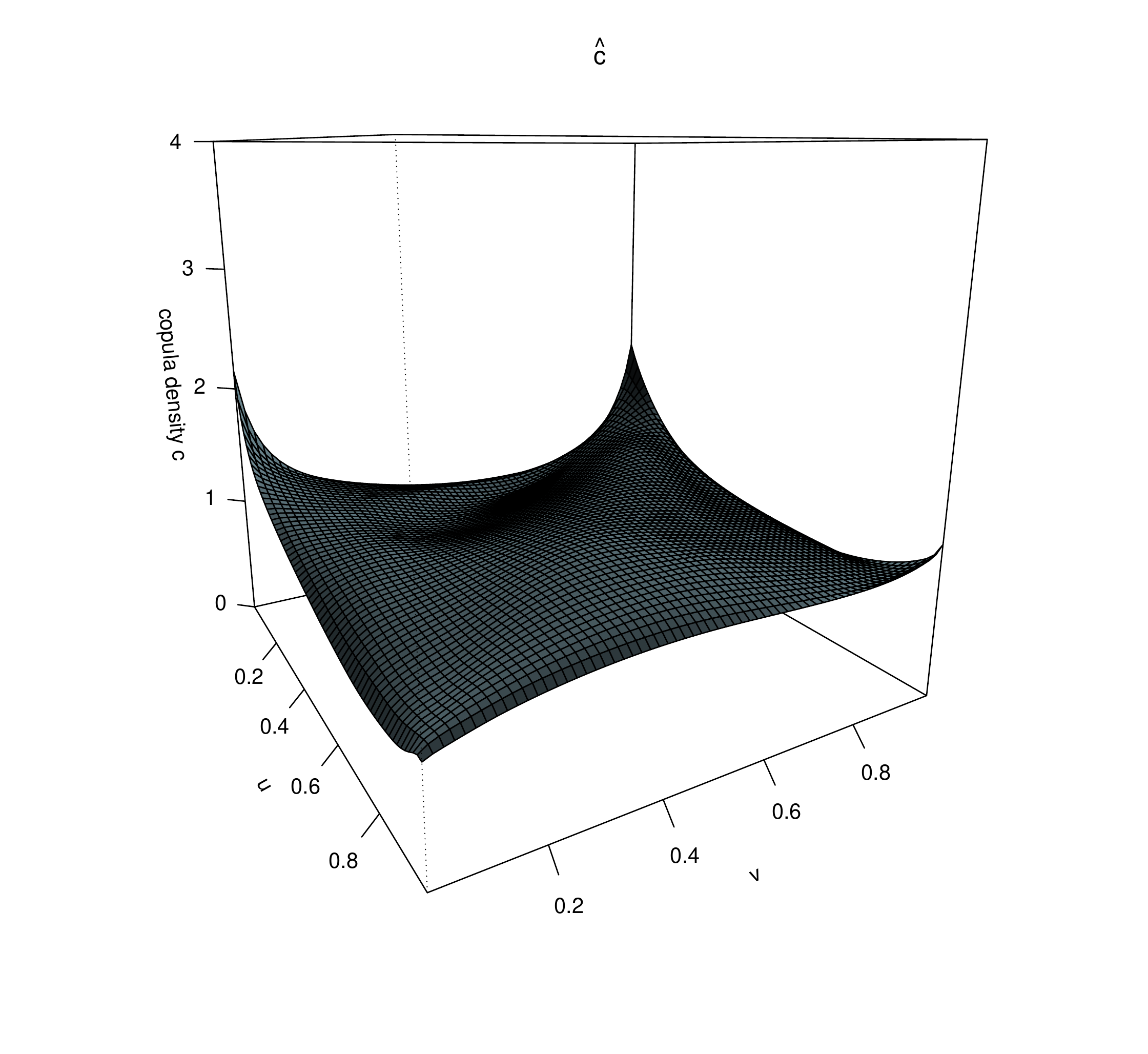}\includegraphics[width=.45\textwidth]{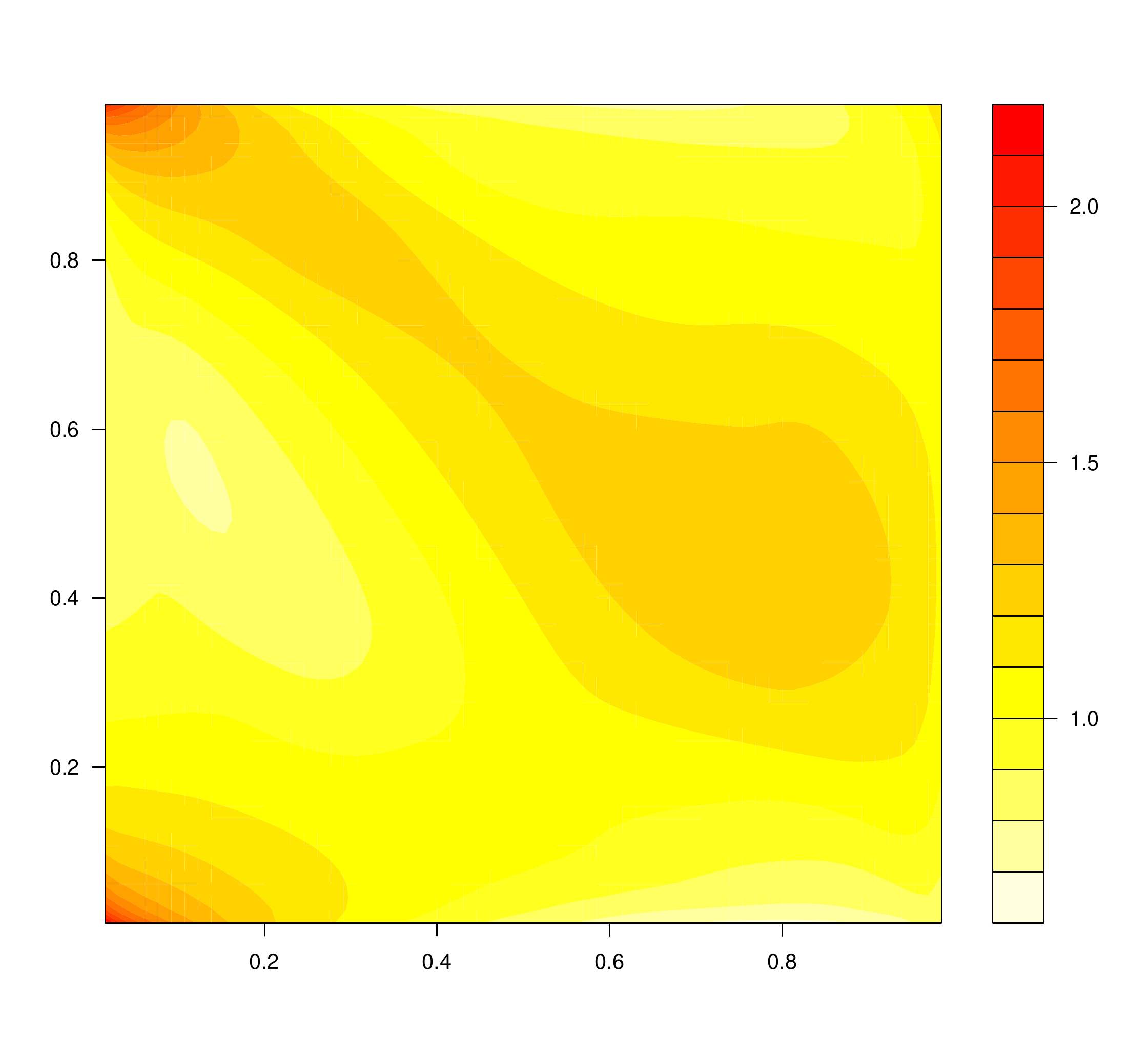}
\caption{Estimated copula density of $(X_{T-1},X_T)$ for the negative log-returns of the S$\&$P500 index.}
\label{fig:copdens}
\end{figure}

\ppn The copula density $c_1$ of $(X_{T-1},X_T)$ is estimated from these data by the LLTKDE2 estimator of \cite{Geenens16}, see Figure \ref{fig:copdens}. The shape of the estimated copula density indicates that $(X_{T-1},X_T)$ is not bivariate Gaussian (Gaussian copulas can only show peaks in opposite corners of $\Is$, not adjacent corners). The `heat' map clearly shows two effects. First, some sort of negative effect, which shows that log-returns on two successive days may be negatively associated. If today's return is low (large value of $u$), one can expect a return tomorrow in the middle range ($v \simeq 0.5$), while if today's return is very high ($u \simeq 0$), chances are that tomorrow's return will be very low (i.e., negative) ($v\simeq 1$). This is, however, largely balanced by the second effect, materialised by the peak in the lower-left corner of the unit square: there is also a substantial probability that, given a high return today ($u \simeq 0$), tomorrow's return is very high as well ($v \simeq 0$). In other words, when $X_{T-1}$ is seen to be small on a day, one expects a more extreme (in either direction) realisation of $X_T$ the day after than in other situations. This, obviously, impacts the corresponding values of $c$VaR in a direct way. 

\begin{figure}[h] \centering
\includegraphics[width=.5\textwidth]{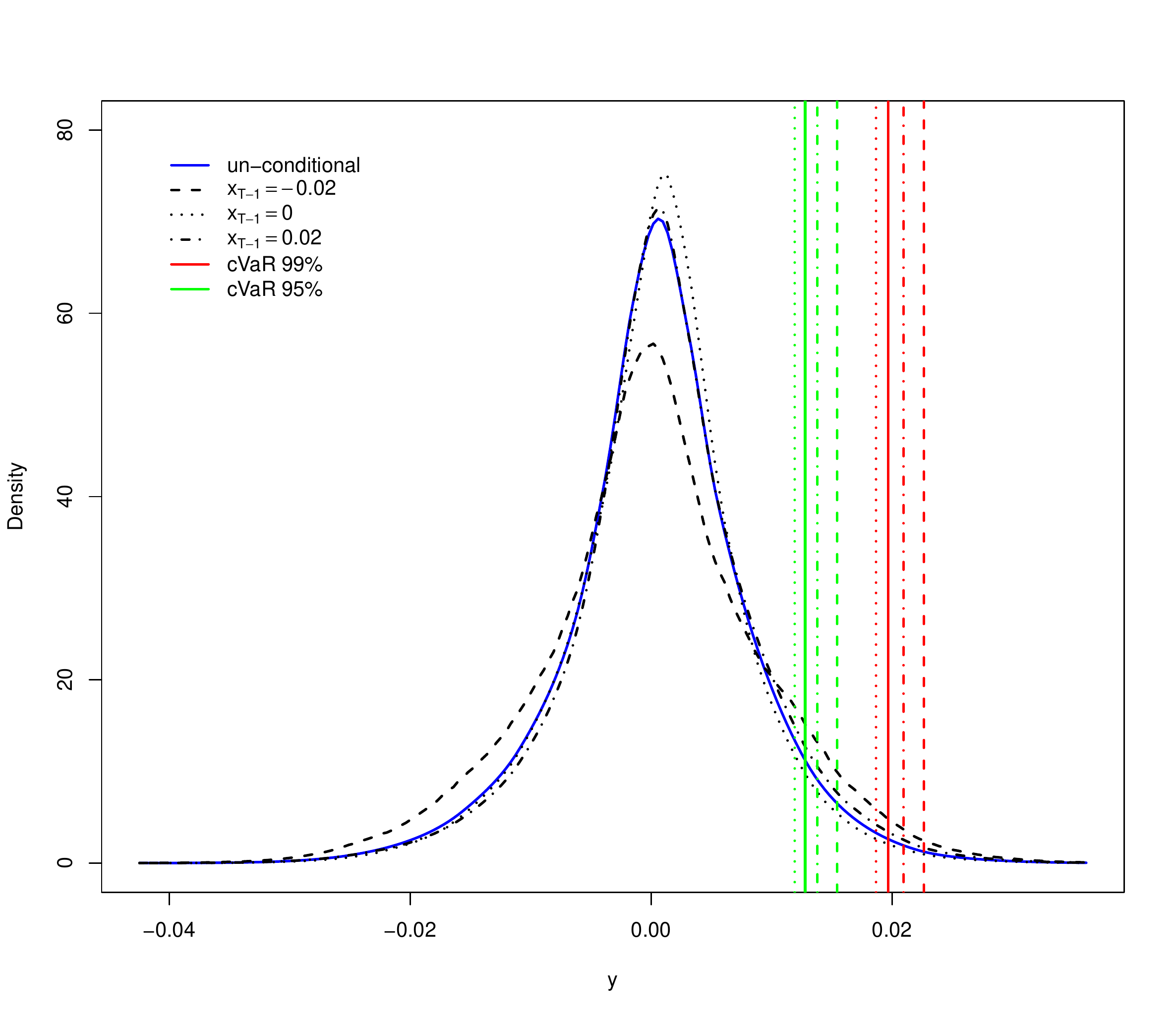}
\caption{Estimated unconditional density $\hat{f}_X$ (plain blue line) of negative log-returns, conditional densities $\tilde{f}_{X_T|X_{T-1}}(\cdot|x)$ for $x = -0.02$ (dashed), $x=0$ (dotted) and $x=0.02$ (dashed-dotted), and corresponding values of $\widehat{\VaR}$ and $c\widehat{\VaR}$ at level 95\% (green) and 99\% (red).}
\label{fig:conddens}
\end{figure}

\ppn This is illustrated by Figure \ref{fig:conddens}. The conditional densities $f_{X_T|X_{T-1}}(\cdot|x)$ for $x = -0.02$ (high return today), $x=0$ (medium return today) and $x=0.02$ (low return today) have been estimated by (\ref{eqn:estconddenscop}). The stationary marginal (i.e., unconditional) density $f_X$ was estimated by the local log-quadratic estimator \citep{Loader96}, and is shown in Figure \ref{fig:conddens} (blue line). For $x = -0.02$, which here corresponds to $u \simeq 0.01$, $\hat{f}_X$ is multiplied on its domain by the `slice' of $\hat{c}_1$ at $u \equiv 0.01$ which is a U-shaped function. Hence the (estimated) conditional density of $X_T$ given $X_{T-1}=-0.02$ has substantially fatter tails. This translates into a higher $c\widehat{\VaR}$ at both levels $\alpha = 0.95$ and $\alpha = 0.99$ (i.e., higher risk). When $x_T = 0$, the risk is actually lower than average ($c\widehat{\VaR}$ lower than the unconditional VaR), because for $u \simeq 0.5$ the `slice' of $\hat{c}_1$ shows a mode around the middle-range and down-weighs the tails of the resulting conditional density. Along similar lines, when $x_T = 0.02$, one finds that $c\VaR$ is slightly higher then the unconditional one. Note that extracting $c$VaR as a quantile of (\ref{eqn:estconddistrcop}) does not require the estimation of the density $f_{X_T|X_{T-1}}$; here, those densities are shown in Figure \ref{fig:conddens} for illustration purpose only. The obtained values of $\widehat{\VaR}$ and $c\widehat{\VaR}$, at $x = -0.02$, $0$ and $0.02$, for $\alpha = 0.95$ and $0.99$, are given in Table \ref{tab:VaR}.

\ppn \begin{table}[h] \centering
 \begin{tabular}{| c | c | c c c | }
  \hline
 & $\widehat{\text{VaR}}_{\alpha,T}$ & & $c\widehat{\text{VaR}}_{\alpha,T}(x)$ & \\
$\alpha$ &  & $x = -0.02$ & $x = 0$ & $x = 0.02$ \\
\hline
0.95 & 0.01278 & 0.01543 & 0.01192 & 0.01379 \\
0.99 & 0.01968 & 0.02263 & 0.01867 & 0.02095 \\
\hline
 \end{tabular}
\caption{Estimated VaR and $c$VaR for the S$\&$P500 data at level $\alpha = 0.95$ and $\alpha = 0.99$.}
\label{tab:VaR}
\end{table}

\subsection{Monte-Carlo simulation} \label{sec:MC}

\begin{figure}[h] \centering
\includegraphics[width=.7\textwidth]{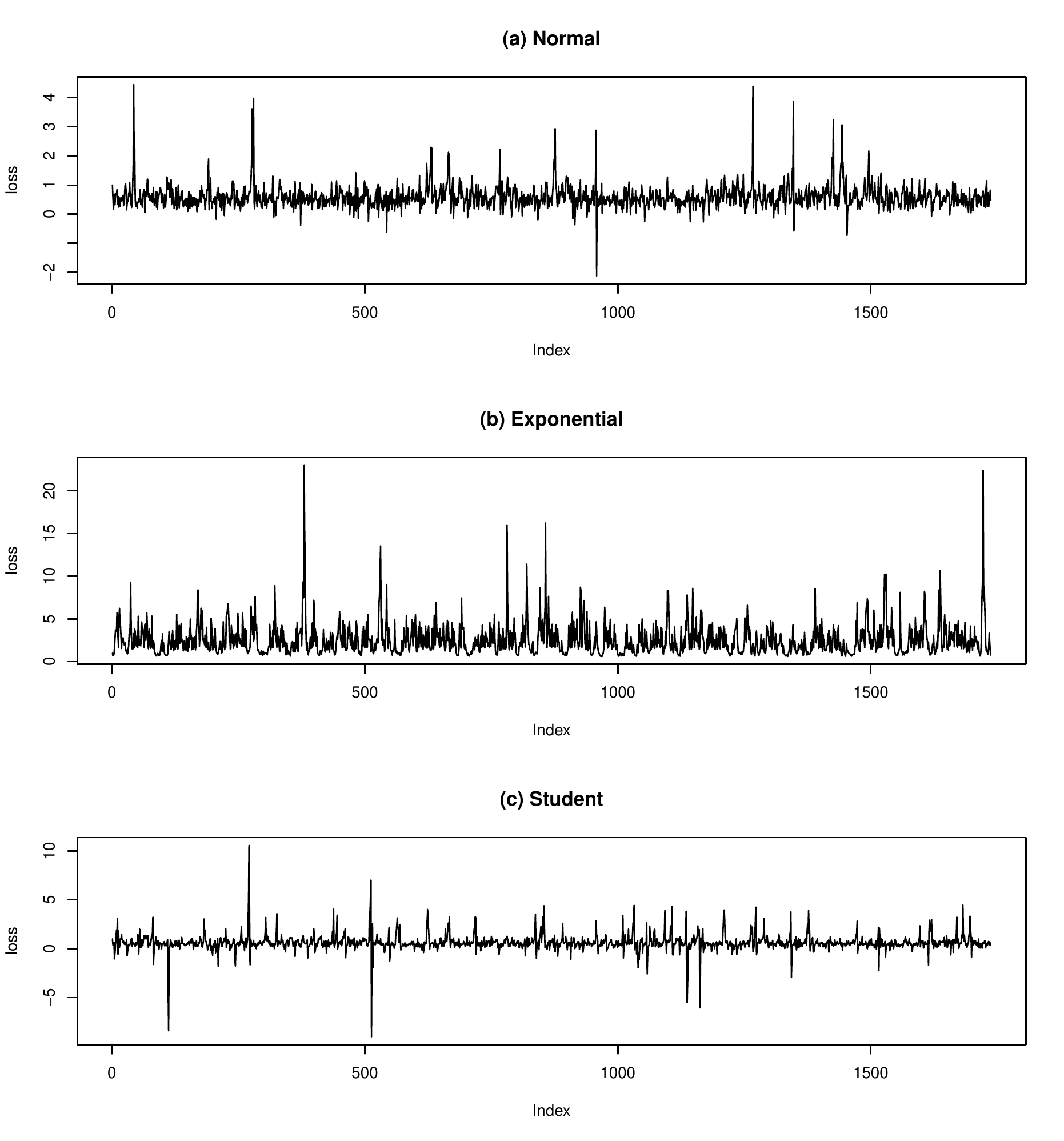}
\caption{Four series of $1,736$ simulated daily losses from (\ref{eqn:MCmod}) with (a) Normal innovations; (b) Exponential innovations; (c) Student-$3$ innovations.}
\label{fig:serMC}
\end{figure}

The previous analysis is, of course, purely descriptive. Here the proposed estimator of $c\VaR$ is evaluated by contrasting its forecasting performance against a range of reasonable alternatives through a Monte Carlo simulation study. Similarly to \citet[Section 3]{Franke15}, consider the nonlinear AR(1)-ARCH(1) model
\begin{equation} X_t = a + b X_{t-1} + \frac{\sqrt{2}}{X_{t-1}} \phi_{c,d}(X_{t-1}) + \sqrt{\omega + \alpha X^2_{t-1}} \ \varepsilon_t, \quad t = 1,2,\ldots, \label{eqn:MCmod} \end{equation}
where $a=0.4$, $b=0.3$, $\phi_{c,d}$ is the normal density with mean $c$ and standard deviation $d$, $c=1.657$, $d=0.1175$, $\omega = 0.007$, $\alpha = 0.2$, and the innovations $\varepsilon_t$ are i.i.d.\ and follow a distribution $\Psi$. Clearly, under this model,
\begin{equation} c\VaR_{\alpha,T}(x) = a + b x + \frac{\sqrt{2}}{x} \phi_{c,d}(x) + \sqrt{\omega + \alpha x^2} \ \Psi^{-1}(\alpha), \label{eqn:trueCVAR} \end{equation}
where $\Psi^{-1}(\alpha)$ is the quantile of level $\alpha$ of the distribution $\Psi$. Four series of $1,736$ `daily' log-returns were generated following (\ref{eqn:MCmod}), seeded from $X_0 = 1$, with (a) $\Psi = $ the standard normal distribution; (b) $\Psi = $ the standard Exponential distribution; and (c) $\Psi=$ the Student-$t_3$ distribution. Together these series mimic stylized features of real financial data, viz.\ volatility clustering, asymmetry and fat tails, see Figure \ref{fig:serMC}.

\ppn On each series, the copula density $c_1$ was first estimated on the first $T=252$ observations, which corresponds to roughly one year of trading days, a common time horizon in rolling window analyses. The conditional $\VaR$ at time $T+1=253$ given $X_T=x_T$ was then estimated via the procedure described in the previous subsection. Then a rolling window of width $N=252$ was used: each day, the oldest observation was discarded and the newly observed one included in the `learning sample', thereby updating on a daily basis the estimation of the copula density for forecasting the next $c\VaR$. This produced, for each of the innovation distributions, two collections of 1,484 one-step-ahead $c$VaR forecasts, one at each of the two levels $\alpha=0.95$ and $\alpha=0.99$. Those can be compared to the true $c$VaR (\ref{eqn:trueCVAR}). For each couple $(\Psi,\alpha)$, the Mean Squared Error of the nonparametric copula-based estimator (hereafter: NP-Cop), viz.\ $\E\left((c\widehat{\VaR}_{\alpha,T}(X_{T-1}) -c\VaR_{\alpha,T}(X_{T-1}) )^2 \right)$, was finally approximated by its empirical counterpart
\[\widehat{\text{MSE}}_\alpha = \frac{1}{1484} \sum_{t=253}^{1736} (c\widehat{\VaR}_{\alpha,t}(X_{t-1}) -c\VaR_{\alpha,t}(X_{t-1}) )^2.  \]
By the same process exactly, one can approximate the Mean Squared Error of $c$VaR forecasts produced by other procedures: the main nonparametric competitor (DKNW) based on inverting the `double kernel' Nadaraya-Watson estimator (\ref{eqn:DKNWconddistr}), as well as a battery of common GARCH-type models: GARCH, iGARCH, eGARCH and gjr-GARCH, all with three different innovation distributions (Normal $\Ns$, Student $\Ss$ and skewed-Student S$\Ss$). These models have been fitted using the R package  {\tt rugarch} \citep{Ghalanos17}. Also included in the study is the Generalised Autoregressive Score (GAS) model, again with the same three innovation distributions, which has been fitted via the R package {\tt GAS} \citep{Ardia16}. Note that the DKNW estimator was fit from the R package {\tt np} \citep{Hayfield08}. All those (approximated) MSE's are shown in Table \ref{tab:MCres}.

\begin{table}
	\centering
\begin{tabular}{||c || c | c || c | c || c | c ||}
	\hline \hline
 & \multicolumn{2}{|c||}{(a)} & \multicolumn{2}{|c||}{(b)} & \multicolumn{2}{|c||}{(c)} \\
 
	&  $\alpha=0.95$  &  $\alpha=0.99$  &  $\alpha=0.95$  &  $\alpha=0.99$ &   $\alpha=0.95$  &  $\alpha=0.99$ \\
	\hline \hline
 NP-Cop&{\bf 0.0427} &0.2326 &  \uline{2.0982} &  \uline{7.1370}  &  \uline{0.2416}  &   \uline{0.7685}\\
 DKNW &0.1592 &0.2492 & 6.1859&  17.3181  &  0.6351   &  1.6881\\
\hline
GARCH-$\Ns$&0.1286 &\uline{0.2011}& \uline{3.1421} &  9.5377 &   0.4174  &   0.8600\\
GARCH-$\Ss$&0.1570 &0.3816& 3.6141 &  7.7965  &  0.5426  &   1.3535\\
GARCH-S$\Ss$&0.1635 &0.4037 &6.4947 & 17.5205 &   0.5359  &   1.3811\\
iGARCH-$\Ns$&0.1595 &0.2717 &3.4752  & 9.6427 &   0.4622   &  0.9101\\
iGARCH-$\Ss$&0.1613 &0.4465 &4.0734  & 9.3564 &   0.5468  &   1.4122\\
iGARCH-S$\Ss$&0.1707 &0.5201& 6.4096 & 17.6876  &  0.5543  &   1.4994\\
eGARCH-$\Ns$&0.7475 &0.7789 &3.8632 & 11.4942  &  0.6154  &   0.7745\\
eGARCH-$\Ss$&0.3960 &0.4679 & 4.6201&  12.1515   & 0.6829   &  0.9165\\
eGARCH-S$\Ss$&0.4143 &0.4754 &6.0764  &12.2205  &  0.7423  &   1.0688\\
gjr-GARCH-$\Ns$&\uline{0.0871} &{\bf 0.1301}& {\bf 1.9685} &  {\bf 5.9310}  &  {\bf 0.2205}  &  {\bf 0.3861}\\
gjr-GARCH-$\Ss$&\uline{0.1074} &\uline{0.1966} & 3.2292  & \uline{6.7643} &   0.3777  &   0.8670\\
gjr-GARCH-S$\Ss$&0.1137 &0.2190& 5.2181&  13.6359  &  \uline{0.3233}  &   \uline{0.7527}\\
GAS-$\Ns$&0.6413 &1.2018 &10.9377 &21.6921 &0.8957 &1.7906\\
GAS-$\Ss$&0.2337 &0.3872 &6.3169 & 12.1580   & 0.5512   &  1.0466\\
GAS-$\Ss$&0.2177 &0.3667 &9.9595 & 21.5550  &  0.5430   &  1.0082\\
\hline \hline
\end{tabular}
\caption{(Approximated) Mean Squared Errors for $c$VaR forecasts at two levels $\alpha = 0.95$ and $\alpha = 0.99$, for three innovation distributions $\Psi$: (a) Normal, (b) Exponential and (c) Student-3. The minimum MSE for each couple $(\alpha,\Psi)$ is in bold, the second and third lowest MSE's are underlined.}
\label{tab:MCres}
\end{table}

\ppn For these series, the best parametric model appears to be the gjr-GARCH with Normal innovations, and this regardless of $\Psi$. Importantly, the nonparametric copula-based is actually right behind in terms of Mean Squared Error, with the lowest MSE in case (a) at level $\alpha = 0.95$ and MSE consistently among the lowest three all across the other scenarios. It seems thus fair to say that the proposed `NP-Cop' estimator is on par with the best parametric models in terms of accuracy of estimation, on nonlinear models like (\ref{eqn:MCmod}). The power of this observation lies in that no model choice is required for the nonparametric estimator, and as such, the risk of model misspecification is evidently null. Indeed it stands out from Table \ref{tab:MCres} that, if it may be {\it possible} to find a parametric model producing $c$VaR forecasts slightly more accurate than the NP-Cop procedure (in this simulation: the gjr-GARCH model), any other choice than that best model typically results in inferior performance. Naturally, when facing real financial data in practice, one usually has very little idea about what could be that best model. Worse, marginally different model specifications may lead to very different results: for instance, in this simulation study, the gjr-GARCH with Skewed-Student innovations produces very bad results under scenario (b), while the same model but with Normal or even Student innovations appears very good. On the other hand, that same gjr-GARCH model with Skewed-Student innovations happens to be a good choice for scenarios (a) and (c). The choice of a parametric model is thus particularly sensitive, which instates the proposed `model-free' nonparametric copula-based procedure as the perfect default choice.

\subsection{Forecasting and Backtesting - IBM data} \label{subsec:IBM}

\begin{figure}[H] \centering
	\includegraphics[width=.7\textwidth]{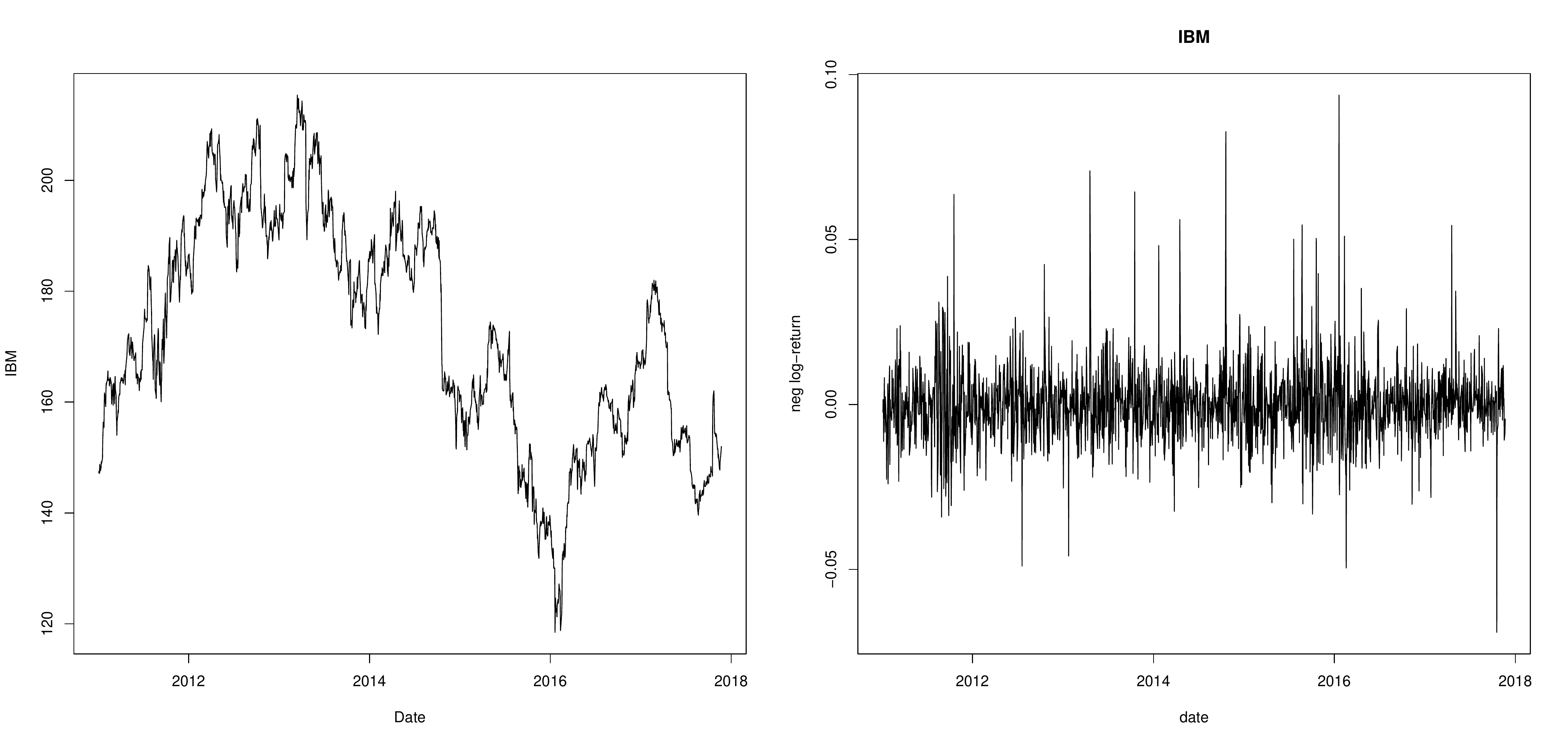}
	\caption{IBM Corporation stock index at opening (left) and corresponding negative log-return series (right) for the period January 2011 to November 2017 (1,736 trading days).}
	\label{fig:IBMdat}
\end{figure}

In order to assess the performance of the NP-Cop procedure under real conditions, the IBM Corporation stock index from the 3rd of January 2011 to the 22nd of November 2017 (1,736 trading days) is now considered. As an individual stock, the IBM data are less `smooth' (in some sense) that aggregated index data such as the S\&P500 index considered above, hence less likely to follow well understood simple parametric specifications. Its modelling is thus more challenging. The raw series, as well as its negative log-returns, are shown in Figure \ref{fig:IBMdat}. 

\begin{figure}[H] \centering
	\includegraphics[width=.9\textwidth]{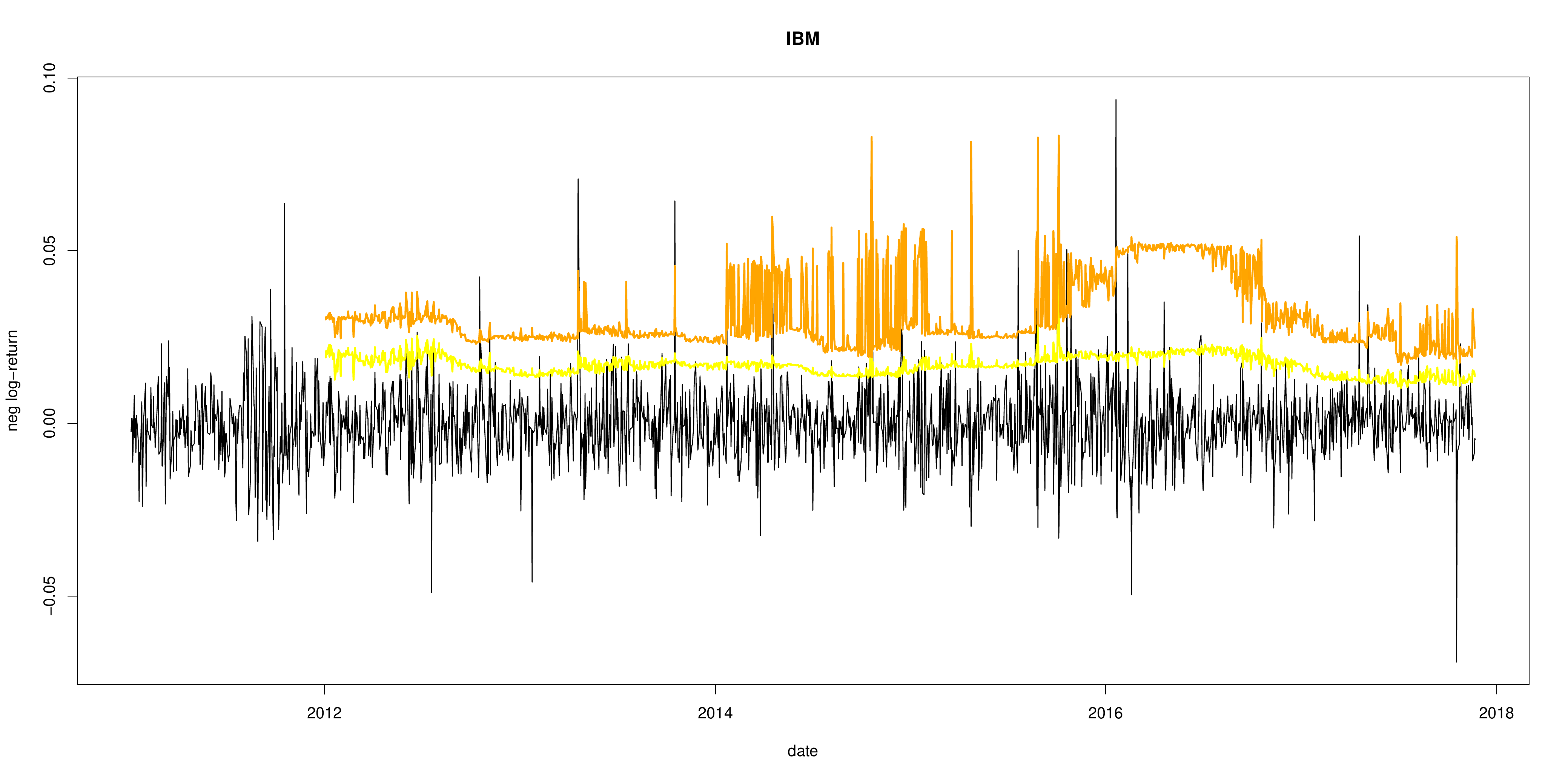}
	\caption{IBM data: daily one-step-ahead $c\VaR$ forecasts at level 95\% (yellow) and 99\% (orange), rolling window of width $N=252$ (roughly one year).}
	\label{fig:IBMVaR}
\end{figure}

\ppn Exactly as in Section \ref{sec:MC}, the copula density $c_1$ of $(X_{T-1},X_T)$ was first estimated from the first year of data (first $T=252$ observations), and the conditional $\VaR$ at time $T+1=253$ given $X_T=x_T$ was then estimated as in Section \ref{subsec:SNP}. Following a rolling window (of width $N=252$) procedure, the current estimate of $c_1$ was then updated on a daily basis for forecasting the next $c\VaR$. The series of daily one-step-ahead $c\VaR$ forecasts at level $95\%$ (yellow) and $99\%$ (orange) are shown in Figure \ref{fig:IBMVaR}. Over the 1,484 $c\VaR$ forecasts, the proportion of $\VaR$ violations (i.e., when the realised value of the loss exceeds the forecast $\VaR$) is $0.0498$ for $c\VaR$ at level 95\% and $0.013$ for $c\VaR$ at level 1\%. Both are very close to their target.

\ppn Many procedures, essentially akin to statistical hypothesis tests, have been proposed for assessing and comparing such $c\VaR$ forecasts on real data. The Basel accords appraise their accuracy through `back-testing', hence the same methodology will be followed here. \citet{Campbell07,Gaglianone11} and \cite{Nieto16} give comprehensive reviews of such back-testing procedures. The simplest version is simply to contrast the empirical proportions of violations against their targeted theoretical probability $1-\alpha$. This is \cite{Kupiec95}'s `unconditional coverage' test (UC). This, however, ignores the likely serial correlation of the violation events. \cite{Christoffersen98} suggested a `conditional coverage' test (CC) taking this into account. Although by far the most popular among practitioners, these two tests have been criticised on different grounds, see \cite{Escanciano10}. The favoured option seems to be the `dynamic quantile' test (DQ) of \cite{Engle06}, which jointly tests for both UC and CC and has proved more powerful against some forms of malfunctions in the forecasting. Another alternative could be the one based on quantile regression (`VQR' test) proposed in \cite{Gaglianone11}. However, that test is known \citep[Section 4.1]{Gaglianone11} to be asymptotically equivalent to the DQ test, so it was not considered further given that the analysed time series here covers more than 6 years of data. 

\begin{table}
	\centering
\begin{tabular}{||c | c c c | c c c||}
	\hline \hline
	& & $\alpha=0.95$ & & & $\alpha=0.99$ & \\
	& UC & CC & DQ & UC & CC & DQ \\
	\hline
	NP-Cop&0.986&0.793&0.219&0.297&0.454&0.106 \\
	DKNW &0.649&0.814&\uline{0.007}&0.580&0.360&\uline{0.011} \\
	\hline
	GARCH-$\Ns$&0.173&0.162&0.075&\uline{0.016}&\uline{0.010}&\uline{0.001} \\
	GARCH-$\Ss$&0.206&0.171&\uline{0.003}&\uline{0.049}&\uline{0.021}&\uline{0.000} \\
	GARCH-S$\Ss$&0.826&0.314&\uline{0.013}&0.580&0.704&0.234 \\
	iGARCH-$\Ns$&0.033&0.090&\uline{0.019}&\uline{0.049}&\uline{0.021}&\uline{0.002} \\
	iGARCH-$\Ss$&0.797&0.422&0.067&0.763&0.802&0.923 \\
	iGARCH-S$\Ss$&0.323&0.155&\uline{0.016}&0.827&0.854&0.912 \\
	eGARCH-$\Ns$&\uline{0.030}&\uline{0.001}&\uline{0.000}&\uline{0.000}&\uline{0.000}&\uline{0.000} \\
	eGARCH-$\Ss$&\uline{0.030}&0.063&0.056&\uline{0.028}&0.060&0.104 \\
	eGARCH-S$\Ss$&0.649&0.900&\uline{0.001}&0.200&0.335&\uline{0.015} \\
	gjr-GARCH-$\Ns$&0.618&0.819&0.076&\uline{0.008}&\uline{0.024}&0.090 \\
	gjr-GARCH-$\Ss$&0.300&0.325&\uline{0.020}&0.297&0.292&\uline{0.043} \\
	gjr-GARCH-S$\Ss$&0.826&0.314&\uline{0.007}&0.423&0.335&\uline{0.021} \\
	GAS-$\Ns$&\uline{0.017}&\uline{0.047}&0.115&\uline{0.049}&0.095&0.062 \\
	GAS-$\Ss$&0.706&0.644&0.160&\uline{0.004}&\uline{0.014}&\uline{0.005} \\
	GAS-$\Ss$&0.458&0.591&0.052&\uline{0.049}&0.095&\uline{0.021} \\
	\hline \hline
\end{tabular}
\caption{$p$-values associated to the three backtesting procedures (UC, CC, DQ) for the NP-Cop procedure and its competitors. Underlined values highlight statistical significance (at 5\% level).}
\label{tab:pval}
\end{table}

\ppn Table \ref{tab:pval} shows the $p$-values associated to the 3 tests (UC, CC, DQ) for the 17 $c\VaR$ forecasting procedures investigated in Section \ref{sec:MC}, at levels 95\% and 99\%. It appears that the adequacy of the forecasts is rejected (at significance 5\%) by at least one test for at least one level for all but two of the tested procedures: the proposed nonparametric copula-based estimator, and the integrated GARCH model with Student innovations. For all other procedures there is some evidence that the $c\VaR$ forecasts are not adequate in some sense. In order to compare further the proposed NP-Cop to the parametric iGARCH-$\Ss$ specification, one resorts to the `Quantile Loss' function \citep{Giacomini05} which, for the forecast at time $t$, is
\[\ell_{\alpha,t} \doteq \ell\left(X_t,c\widehat{\VaR}_{\alpha,t}(x_{t-1})\right) = (\alpha-\indic{X_t \leq c\widehat{\VaR}_{\alpha,t}(x_{t-1})})(X_t-c\widehat{\VaR}_{\alpha,t}(x_{t-1})).\]
This is obviously an asymmetric loss function which penalises more heavily Value-at-Risk exceedance than otherwise \citep{Gonzalez04}. Averaging over the 1,484 predictions, one obtains for the nonparametric copula-based procedure losses of
\[\frac{1}{1484} \sum_{t=253}^{1736} \ell_{0.95,t} = 0.00139 \qquad \text{ and } \qquad \frac{1}{1484} \sum_{t=253}^{1736} \ell_{0.99,t} = 0.00059, \]
while for the parametric iGARCH-$\Ss$ model, one gets
\[\frac{1}{1484} \sum_{t=253}^{1736} \ell_{0.95,t} = 0.00139 \qquad \text{ and } \qquad \frac{1}{1484} \sum_{t=253}^{1736} \ell_{0.99,t} = 0.00055. \]
On this criterion, one can say that both procedures are doing equally well. It is thus fair to conclude that the proposed nonparametric copula-based procedure is on par with {\it the best} parametric procedure when it comes to forecast the one-step ahead $c\VaR$ at level 95\% and 99\% for the IBM data. This is impressive, given that this was achieved without any parametric guidelines, and rather extracting the relevant information straight from the data.

\begin{figure}[H] \centering
	\includegraphics[width=.9\textwidth]{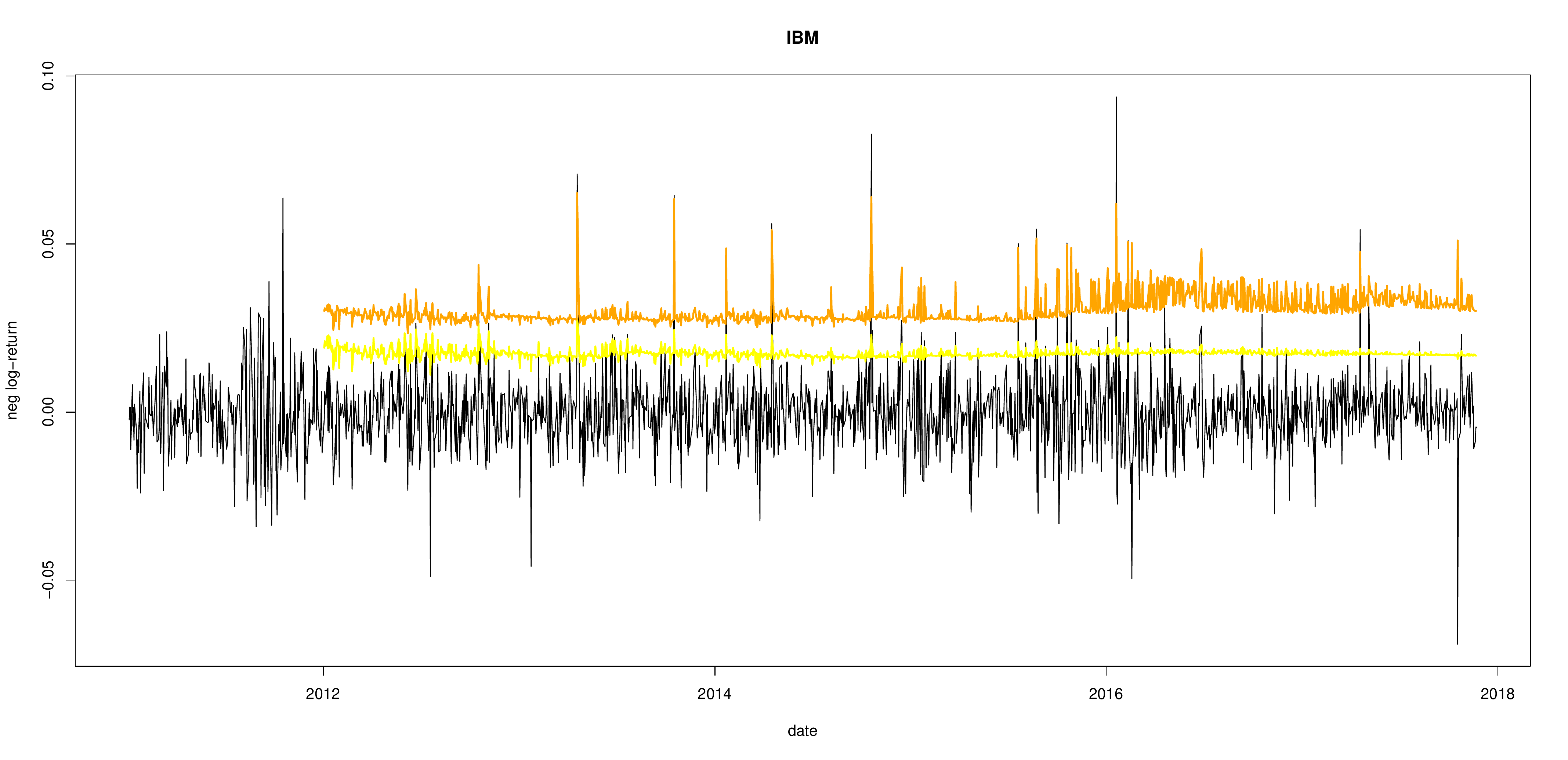}
	\caption{IBM data: daily one-step-ahead $c\VaR$ forecasts at level 95\% (yellow) and 99\% (orange), expanding window.}
	\label{fig:IBMVaR2}
\end{figure}

\ppn Instead of using a rolling window as previously, one can also use an {\it expanding} window, that is, one keeps all the previous observations from time 0 when updating the estimate of the copula density from the newly observed returns. This is, obviously, meaningful if one believes in the stationarity of the series over the whole period of observation. If one does so on the IBM series, one obtains the daily one-step ahead $c\VaR$ forecasts shown in Figure \ref{fig:IBMVaR2}.

\ppn The series of $c\VaR$ are naturally smoother than in the `rolling' window case (especially at the 99\%-level), as one can understand. What is remarkable, though, is that the procedure seems able to guess each time there is going to be an extreme loss: each time the realised series shows a peak higher than, say, 0.04, the estimated $c\VaR$ at level 99\% shows a peak of similar amplitude as well. This seems very promising and motivates further study of the proposed procedure.
  
\section{Concluding remarks and future work} \label{sec:ccl}

This paper investigates a novel nonparametric conditional Value-at-Risk forecast procedure. Like previously suggested nonparametric methods, the $\VaR$ is here obtained by direct inversion of a nonparametric estimator of the conditional cumulative distribution of interest. What differs from them is that estimation of that conditional distribution is based on the density of the copula describing the dynamic dependence observed in the series of returns. This has several advantages, as expounded in the paper. In particular, the so-produced estimated conditional distribution function is always constrained to between 0 and 1 and increasing, which makes it easy to invert for extracting the desired quantiles. The copula framework leads to intuitive interpretation of the results, see application on the S\&P500 index in Section \ref{subsec:SNP}. In addition, Monte-Carlo simulations (Section \ref{sec:MC}) and the analysis of the IBM Corporation stock from January 2011 to November 2017 (Section \ref{subsec:IBM}) have revealed that the suggested procedure may perform as well as {\it the best} parametric models for one-step ahead Value-at-Risk forecasting. 

\ppn It is important to note that $\VaR$ has recently been the object of discussion and criticism, mainly because it may be not sub-additive and non-coherent for heavy-tailed loss distributions, see discussion in \citet{Danielsson05} and \citet{Ibragimov07}. Hence other risk measures have been proposed as well, making the choice of the right risk measure a problem of theoretical interest of its own \citep{Cherubini12}. In any case, the Basel III accords recommend complementing the $\VaR$ with the Expected Shortfall (ES), owing to the guaranteed coherence of the latter. Hence estimation methods for the {\it conditional Expected Shortfall} have been suggested and investigated  \citep{Scaillet05,Cai08,Chen08,Kato12,Linton13,Xu16}; see \cite{Nadarajah14} for a review. For some level $\alpha$, the (conditional) ES is defined as
\[c\text{ES}_{\alpha,T}(x) = \E\left(X_T|X_T > c\VaR_{\alpha,T}(x),X_{T-1}=x \right), \]
i.e., the expected loss given that the loss exceeds the corresponding (conditional) Value-at-Risk, that is, 
\[c\text{ES}_{\alpha,T}(x) = \frac{1}{1-\alpha} \int_{c\VaR_{\alpha,T}(x)}^\infty y f_{X_{T}|X_{T-1}}(y|x)\,dy. \]
Define $W_T = X_T -c\VaR_{\alpha,T}$, and see that 
\[c\text{ES}_{\alpha,T}(x) = c\VaR_{T,\alpha}(x) + \frac{1}{1-\alpha} \int_{0}^\infty w f_{W_{T}|X_{T-1}}(w|x)\,dw. \]
From (\ref{eqn:conddenscop}), one has
\[f_{W_{T}|X_{T-1}}(w|x) = f_{W}(w) \times c^{(W)}_{1}(F_X(x),F_W(w)), \]
where $c^{(W)}_1$ is the copula density of the vector $(X_{T-1},W_T)$. Now, because $W_T$ is just $X_T$ plus a constant, the dependence within $(X_{T-1}, W_T)$ is the exact same as within $(X_{T-1}, X_T)$, and their copula is the same. Hence, the conditional Expected Shortfall can be written
\[c\text{ES}_{\alpha,T}(x) = c\VaR_{\alpha,T}(x) + \frac{1}{1-\alpha} \int_{0}^\infty w c_1(F_X(x),F_X(w+c\VaR_{\alpha,T}(x)))f_W(w)\,dw. \]
This, in turn, suggests a nonparametric estimator of type
\[c\widehat{\text{ES}}_{\alpha,T}(x) = c\widehat{\VaR}_{\alpha,T}(x) + \frac{1}{1-\alpha} \frac{1}{T} \sum_{t=0}^{T-1} (X_t-c\widehat{\VaR}_{\alpha,T}(x)) \hat{c}_1(\widehat{F}_X(x),\widehat{F}_X(X_t))\indic{X_t>c\widehat{\VaR}_{\alpha,T}(x)}, \]
with the same estimators as in (\ref{eqn:estconddistrcop}). This estimator will be studied in detail in a forthcoming paper.

\section*{Acknowledgements}
This research was supported by a Faculty Research Grant from the Faculty of Science, UNSW Sydney (Australia).

\end{document}